\documentclass[aps,prb,twocolumn,letterpaper,superscriptaddress,showpacs]{revtex4-2}
\usepackage{graphicx}
\usepackage{CJK}
\usepackage{verbatim}
\usepackage{physics}
\usepackage[colorlinks,linkcolor=blue,anchorcolor=blue, citecolor=blue,urlcolor=blue,]{hyperref}
\usepackage{lineno} 
\usepackage{bm}
\usepackage{mathptmx}
\makeatletter

\begin{document}
\begin{CJK*}{UTF8}{bsmi}
\title{Variation of carrier density in semimetals via short-range correlation:\\
A case study with nickelate NdNiO$_2$}
\author{Ruoshi Jiang (\CJKfamily{gbsn}姜若诗)}
\affiliation{Tsung-Dao Lee Institute \& School of Physics and Astronomy, Shanghai Jiao Tong University, Shanghai 200240, China}
\author{Zi-Jian Lang (\CJKfamily{gbsn}郎子健)}
\affiliation{Tsung-Dao Lee Institute \& School of Physics and Astronomy, Shanghai Jiao Tong University, Shanghai 200240, China}
\author{Tom Berlijn}
\affiliation{Center for Nanophase Materials Sciences, Oak Ridge National Laboratory, Oak Ridge, Tennessee 37831, USA}
\author{Wei Ku (\CJKfamily{bsmi}顧威)}
\altaffiliation{corresponding email: weiku@sjtu.edu.cn}
\affiliation{Tsung-Dao Lee Institute \& School of Physics and Astronomy, Shanghai Jiao Tong University, Shanghai 200240, China}
\affiliation{Key Laboratory of Artificial Structures and Quantum Control (Ministry of Education), Shanghai 200240, China}
\affiliation{Shanghai Branch, Hefei National Laboratory, Shanghai 201315, People's Republic of China}
\date{\today}

\begin{abstract}
Carrier density is one of the key controlling factors of material properties, particularly in controlling the essential correlations in strongly correlated materials.
Typically, carrier density is externally tuned by doping or gating, and remains fixed below room temperature.
Strangely, the carrier density in correlated semimetals is often found to vary sensitively against weak external controls such as temperature, magnetic field, and pressure.
Here, we develop a realistic simulation scheme that incorporates interatomic \textit{noncollinear} magnetic correlation without a long-range order.
Using the recently discovered nickelate superconductor as an example, we demonstrate a rather generic low-energy mechanism that in semimetals short-range correlation can \textit{reversely} modulate the carrier density as well.
Such a mutual influence between correlation and carrier density provides an extra ingredient for sensitive bifurcating behavior.
This special feature of correlated semimetals explains their versatile carrier density at low energy and opens up new possibilities of functionalizing these materials.

\end{abstract}
\maketitle
\end{CJK*}

\section{INTRODUCTION}
The effect of electronic correlation on physical properties of strongly correlated materials is one of the most important topics in condensed matter physics.
Unlike the kinetic energy which dominates the low-energy physics of weak interacting systems, correlation between electrons in strongly correlated materials can introduce significant complexity, leading to the emergence of numerous phenomena, such as the interaction-driven metal-insulator transition~\cite{Mott1949, Mott1968}, colossal magnetoresistivity~\cite{Jonker1950}, unconventional superconductivity~\cite{Steglich1979, Norman2011}, strange metallicity~\cite{Varma1989}, bad metal behavior~\cite{Fisk1976}, quantum spin liquid realization~\cite{Anderson1973}, etc.
Naturally, the most essential quests of the field are centered around exploring these complex correlation effects and efficient means of their control for practical applications.

Among the key controlling factors of electronic correlation, carrier density is known to be the most effective.
This naturally follows the fact that correlation results from influence of electrons onto each other and is therefore sensitive to their average distance, or their density.
Indeed, one typically finds a rich phase diagram in correlated materials hosting dramatically different behaviors upon tuning the carrier density in these materials~\cite{Dagotto1994,Dagotto2005}.
Well-known examples include cuprates~\cite{Taillefer2010}, iron pnictides~\cite{Zhao2008}, manganites~\cite{Schiffer1995}, titenates~\cite{Imada1998}, ruthenates~\cite{Nakatsuji2000}, cobaltates~\cite{Foo2004}, and twisted bilayer graphene~\cite{Cao2018_1,Cao2018_2}, all testifying the extreme efficiency of carrier density in tuning the correlation and in turns the physical properties.

In correlated semimetals (semiconductors with a negative band gap), this strong sensitivity to carrier density grands extra complexity and functionality due to semimetals' additional flexibility in carrier density.
Unlike regular metals and doped semiconductors that have rather robust carrier densities (roughly speaking the size of the Fermi pockets) fixed by their chemical potentials, semimetals have the additional freedom in varying simultaneously the densities of the coexisting electron and hole carriers $n_e$ and $n_h$, since the chemical potential only pins the total electron count, which depends on the \textit{difference} between them, $n_e - n_h$.
Indeed, for example in unconventional high-temperature superconductors such as FeSe~\cite{Kawai2018} and nickelates~\cite{Li2019}, the observed strong temperature dependent Hall coefficients suggests a sensitively varying carrier density.
Such variation of carrier density is in good consistency with the apparent change of size of the Fermi surface observed in FeSe~\cite{Kushnireko2017} by angular-resolved photoemission spectroscopy.
As another example, in unconventional WTe$_2$ superconductor the Hall coefficient~\cite{Kang2015} displays a strong pressure dependence, and the carrier density changes balance over temperature as well~\cite{Pan2017}.
These examples demonstrate the intimate connection between the rich physical behaviors and the versatile carrier density in correlated semimetal.

Therefore, two essential generic questions concerning the carrier density in correlated semimetals are (1) what key factors control the carrier density variation in these systems, and (2) how the carrier density is able to vary so efficiently against change of ``weak''  (or low-energy) external conditions, for example, temperature, pressure, or magnetic field.
The strongest correlation due to strong intraatomic repulsion is known to be able to enhance the effective mass of carriers~\cite{Nayak2017,Nekrasov2018}.
However, the large energy scale of the local repulsion dictates that such mass enhancement is rather robust and thus unable to vary sensitively against weak external conditions.
Apparently, one needs to seek the answer in the physics of a much lower energy scale, such as interatomic correlations.

However, incorporation of interatomic correlation poses a serious technical challenge to our current theoretical and computational capability.
Perturbation treatments such as the $GW$ approximation~\cite{Hirayama2022,Wei2002_GW} can typically capture long-wavelength physics such as long-range screening, but is inadequate for strong short-range correlations.
On the other hand, the state-of-the-art dynamical mean-field treatment (DMFT)~\cite{Georges1996} only includes the high-energy intraatomic correlation, but ignores the multiple scattering associated with interatomic correlations.

Here, to address the above scientific questions, we develop a density functional theory~\cite{DFT1, DFT2} (DFT)-based computational simulation scheme to reveal the unexplored physical effects of the interatomic correlation in real materials.
Specifically, using NdNiO$_2$ as a prototypical example, we demonstrate strong impacts of \textit{noncollinear} magnetic correlation on the one-body propagator in the \textit{absence} of long-range order.
In addition to nontrivial modification of the band dispersion and the quasiparticle lifetime, we find a clear systematic trend that short-range correlation in semimetals can \textit{reversely} modulate the carrier density.
Such a mutual influence between correlation and carrier density unique in correlated semimetals provides an extra ingredient for sensitive bifurcating behavior of materials.
Particularly, given the sub-eV energy scale of nonlocal correlations, this generic mechanism not only can explain the observed strong carrier density modulation in many correlated semimetals, but also opens up new possibilities of functionalizing these materials via weak external tuning such as temperature, pressure and applied field.

We start by developing the following computational simulation scheme for real materials.
First, using the prototypical parent compound of the nickelate superconductor NdNiO$_2$ as an example, we extract a realistic SU(2)-symmetric high-energy many-body Hamiltonian from DFT calculations.
Second, within self-consistent Hartree-Fock treatment, we compute the electronic one-body spectral function under randomly chosen \textit{unordered} magnetic configurations that contains various constrained \textit{noncollinear} spin directions with \textit{negligible} long-range order parameters.
Finally, we analyze the systematic trend of effects short-range correlation induces on the one-body spectral function by averaging results from magnetic configurations of similar correlation strength.

\section{REALISTIC INTERACTING HAMILTONIAN}
Specifically, to account for various noncollinear magnetic correlation, it is necessary for us to obtain a fully SU(2)-symmetric many-body interacting Hamiltonian in the basis of symmetry-respecting atomiclike Wannier orbitals~\cite{supplementary,Wei2002, Marzari1997}.
We therefore demand that it must be able to \textit{simultaneously} reproduce the band structures from DFT calculations under various magnetic structures, including the fully ferromagnetic (FM) and fully antiferromagnetic (AFM) ones.
Furthermore, since this compound contains partially occupied Nd $f$ orbitals and Ni $d$ orbitals, approximations like the ``LDA+$U$''~\cite{Anisimov1993, Liechtenstein1995, supplementary} or hybrid functionals~\cite{Becke1993} are necessary in the DFT calculations in order to ensure a realistic density.
To this end, we apply the (LDA+$U$)+(many-body) procedure~\cite{lang2021,lang2022} to extract the realistic many-body Hamiltonian~\cite{supplementary}.

Furthermore, owing to the very large intraatomic Coulomb repulsion among the Nd $f$ orbitals, their charge fluctuation in the low-energy sector can be safely ``integrated out'' (e.g. via a Schrieffer-Wolff transformation~\cite{Schrieffer, Zaanen1988}), leaving only their $\frac{3}{2}$ spin degrees of freedom~\cite{Choi2020_1} and their FM coupling to the Nd $d$ orbitals.
The resulting SU(2)-symmetric effective Hamiltonian contains the following leading contributions:
\begin{equation}
    \begin{split}
        H^{\text{eff}} &= \sum_{i,m,\nu}\epsilon_mc^\dagger_{im\nu}c_{im\nu}+\sum_{i,i^\prime,m,m^\prime,\nu}t_{ii^\prime mm ^\prime}c^\dagger_{im\nu}c_{i^\prime m^\prime \nu}\\
        &+\frac{1}{2}\sum_{i,m,m^\prime,m^{\prime\prime},m^{\prime\prime\prime},\nu,\nu^\prime} U_{m m^{\prime\prime}  m^\prime m^{\prime\prime\prime}} c^\dagger_{im\nu}c^\dagger
        _{im^{\prime\prime}\nu^\prime}c_{im^{\prime\prime\prime}\nu^\prime}c_{im^\prime\nu}\\
        &-\frac{1}{2}\sum_{im,\nu,\nu^\prime} J_m\mathbf{S}_i^f\cdot c^\dagger_{im\nu}\bm{\sigma}_{\nu\nu^\prime}c_{i m \nu^\prime},
    \end{split}
    \label{eq1}
\end{equation}
including one-body orbital energy $\epsilon$ and hopping strength $t$ of all orbitals, and intraatomic two-body interaction $U$ among the Ni $d$ orbitals, denoted by creation $c^\dagger_{im\nu}$ and annihilation $c_{im\nu}$ operators of orbitals $m$ and spin $\nu$ within unit cell $i$~\cite{supplementary}.
As mentioned above, Nd $f$ orbitals contribute mainly through their $\frac{3}{2}$ spin $\mathbf{S}_i^f$ degrees of freedom via their FM coupling $J$ to the Nd $d$ orbitals in Eq.~(\ref{eq1}).
Here $\bm{\sigma}$ denotes the usual vector of Pauli matrices.

\begin{figure}
\centering
\includegraphics[width=1\columnwidth]{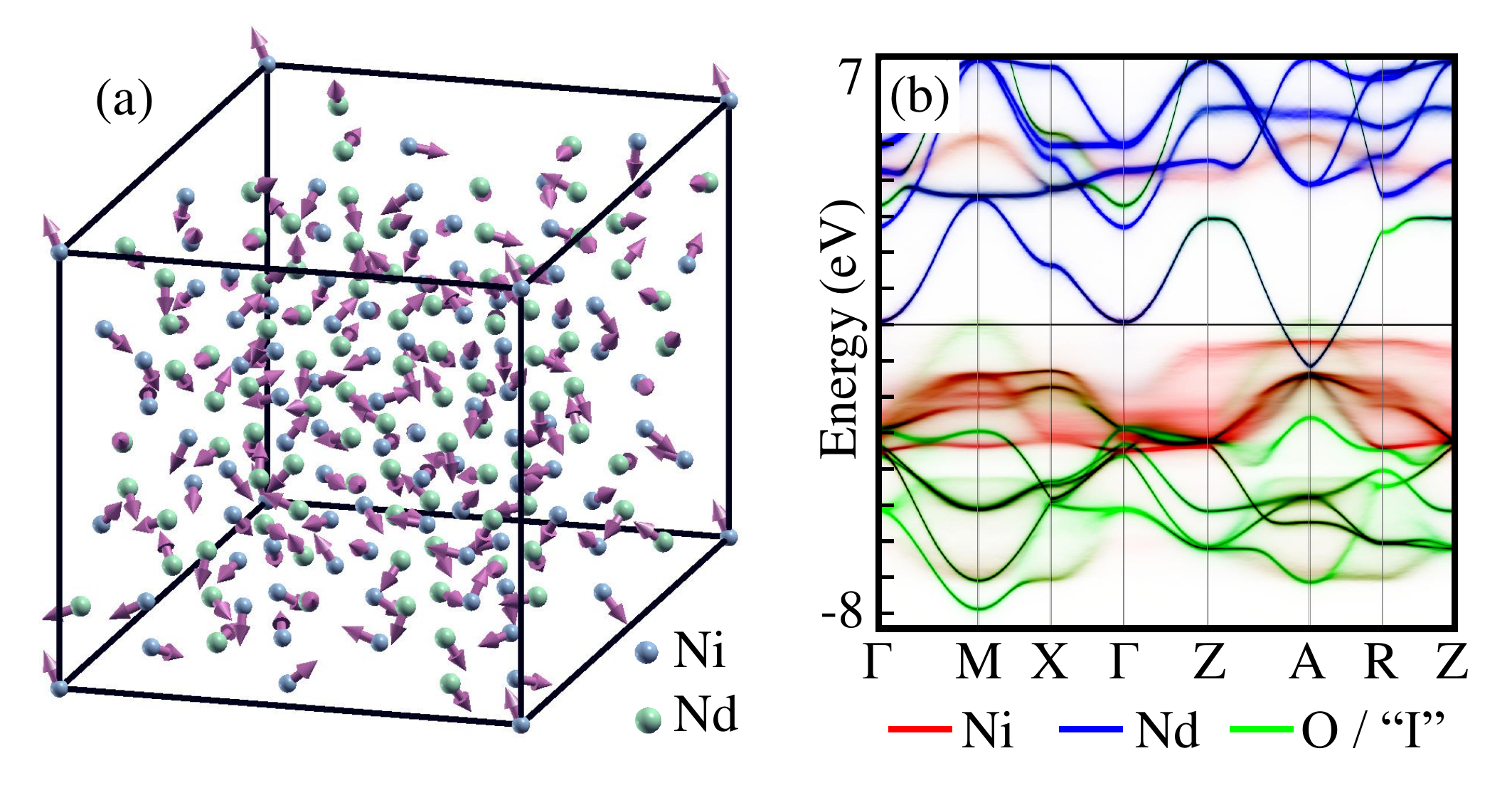}
\vspace{-0.8cm}
\caption{(a) Illustration of \textit{unordered noncollinear} magnetic configurations with spin directions of each magnetic atom denoted by the arrows.
(b) An example of ensemble-averaged one-body spectral function $A(k,m,\omega)$ corresponding to fully Curie-paramagnetic systems with random interatomic correlation (as in the high temperature limit.)
Contributions from different orbitals, including the interstitial ``I'' orbital~\cite{supplementary}, are represented by different colors.
For clarify, Nd $f$ bands are not shown (see text).
}
\vspace{-0.5cm}
\label{fig1}
\end{figure}

\begin{figure*}
\centering
\vspace{-0.5cm}
\includegraphics[width=1\textwidth]{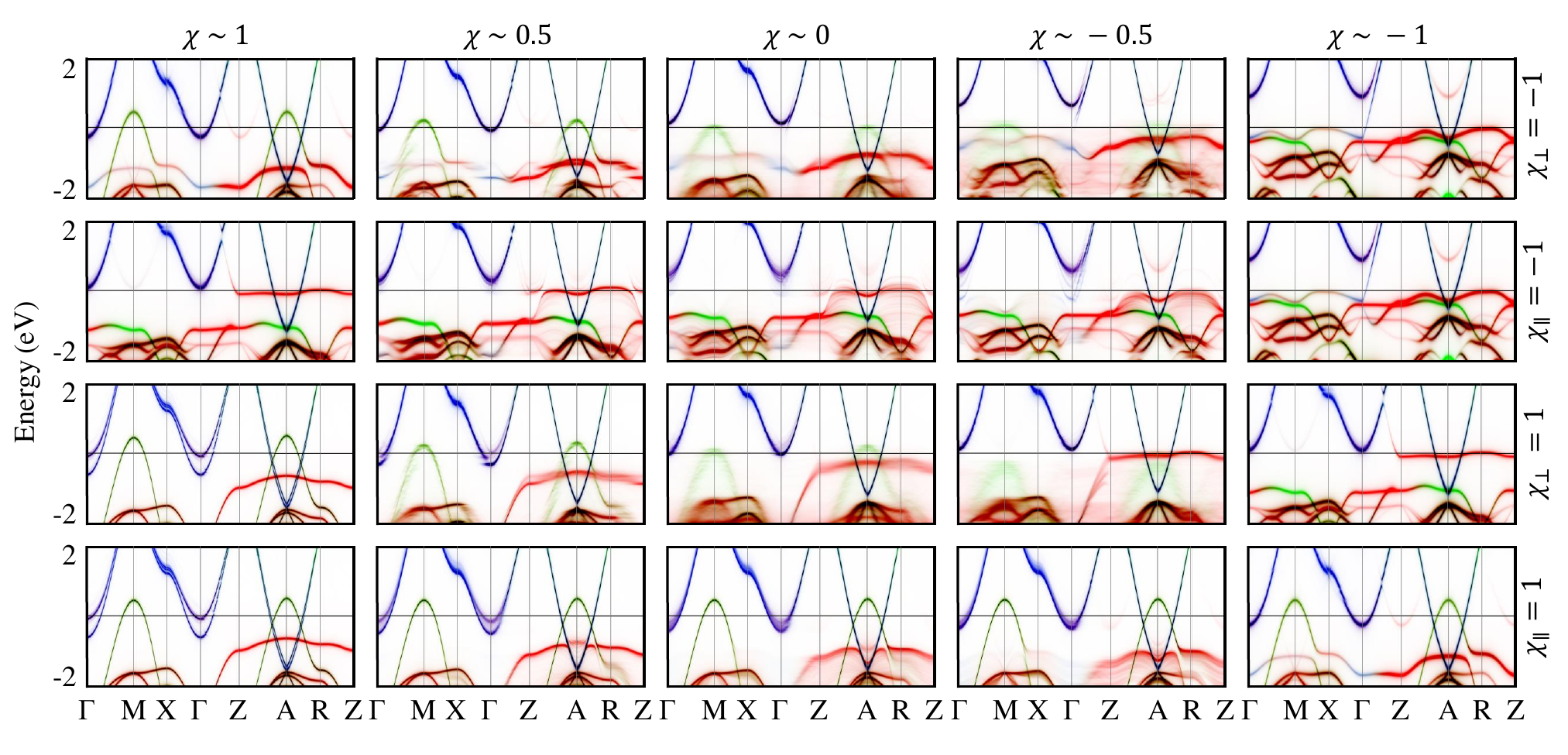}
\vspace{-0.9cm}
\caption{Ensemble-averaged one-body spectral functions $A(k,m,\omega)$ under various nearest neighboring \textit{noncollinear} intralayer correlation $\chi_\parallel$ and interlayer  correlation $\chi_\perp$, while keeping the other direction fixed as AFM (top two rows) or FM (bottom two rows).
The color scheme is the same as in Fig.~\ref{fig1}.
}
\vspace{-0.5cm}
\label{fig2}
\end{figure*}

\section{SIMULATING SHORT-RANGE CORRELATION}
Next, to simulate the impact of short-range correlation on the one-body propagator \textit{in the absence of} long-range order, we construct many randomly oriented supercells of various shapes (typically containing 400 atoms) representing noncollinear magnetic configurations of Ni and Nd atoms [cf. Fig.~\ref{fig1}(a)] and demand that the average order parameters (FM or AFM) be negligibly small~\cite{supplementary}.
(For the one-body propagator that encapsulates the band dispersion of interest here, such simulation is conceptually clean~\cite{supplementary}.)
[In addition, since the interatomic Nd-Nd and Nd-Ni magnetic couplings are negligibly small~\cite{lang2021} ($<0.1$meV), the directions of Nd spins are treated as completely random in our simulation.]
Obviously, the size of the supercells must be sufficiently large in such simulations to allow configurations with strong short-range correlation and yet negligible long-range order at the same time~\cite{supplementary}.
Such a large supercell size, together with the need for constraining the spin directions, seriously limits our options of affordable many-body methods.

We therefore employ the self-consistent Hartree-Fock approximation to compute the frequency $\omega$-dependent, orbital $M$-projected one-body spectral function,
$A(K,M,\omega) = \sum_J |\braket{KM}{KJ}|^2 \delta(\omega-\epsilon_{KJ})$,
using the eigenvalue $\epsilon_{KJ}$ and eigenvector $\ket{KJ}$ corresponding to band $J$ and crystal momentum $K$ of each supercell, with spin directions \textit{constrained} according to the magnetic configuration.
The use of Hartree-Fock approximation makes affordable the large system size necessary for our study.
Furthermore, since in this particular system Ni is predominantly Ni$^+$ ($d^9$ with one hole), atomic many-body multiplets should be well approximated via Hartree-Fock treatment.

Finally, we categorize the configurations based on the strength of their nearest neighboring (NN) magnetic correlation,
$\chi = \langle\mathbf{S}_{i} \cdot \mathbf{S}_{i^\prime}\rangle\vert_{i^\prime\in \mathrm{NN}(i)}$,
between Ni $\frac{1}{2}$ spins $\mathbf{S}_{i}$ and average the one-body spectral functions within each category.
This average can be easily performed in the configuration-independent orbital basis (momentum $k$ and orbital $m$) of the chemical formula unit, $A(k,m,\omega) = \sum_{K,M} |\braket{km}{KM}|^2 A(K,M,\omega)$, through the unfolding procedure~\cite{Wei2010}, which also facilitates an easier visualization in the standard Brillouin zone.
The chemical potential is then determined from the \textit{averaged} one-body spectral function based on the total occupation of these orbitals.

\section{CURIE PARAMAGNETIC PHASE}
Figure~\ref{fig1}(b) gives an example of the resulting ensemble-averaged one-body spectral function that corresponds to fully Curie-paramagnetic systems with random short-range correlation $\chi\sim 0$ (as in the high temperature limit).
Notice that it displays many interesting characteristics distinct from results of typical non-magnetic calculations.
For example, one observes significant broadening and smearing in some of the bands with Ni $d$ orbitals (in red), reflecting a shorter mean-free path and lifetime of quasiparticles corresponding to these bands.
This is evidently from strong scattering against the unordered Ni $\frac{1}{2}$ spins, since particles in the Ni $d$-shell would experience a strong spin dependent self-energy that varies by a large scale of $U\sim 7$ eV in Eq.~(\ref{eq1}).
In comparison, the scattering of Nd $d$ orbitals (in blue) against the Nd $f$ $\frac{3}{2}$ spins is obviously much less effective due to the rather small $f$-$d$ spin coupling $J\sim0.3$ eV in Eq.~(\ref{eq1}).

Notice that the effects of scattering in our result are strongly momentum dependent.
For example, around -2 eV the band between $\Gamma$ and Z point becomes very broad, while at the same energy the bands around the M point remain sharp.
This is in great contrast to the momentum-independent smearing obtained from the DMFT~\cite{Petocchi2020, Ryee2020,Kitatani2020,Lechermann2021,supplementary}, whose self-energy is strictly intraatomic only.
Our approach on the other hand accounts for the \textit{interatomic} self-energy associated with the essential short-range correlation of interest in this study.

\section{EFFECTS OF SHORT-RANGE CORRELATION}
Figure~\ref{fig2} summarizes our main result, which shows a very strong impact of short-range magnetic correlation on the obtained one-body spectral functions.
Even the eV-scale band dispersion can be dramatically modified by the varying interatomic correlation.
Specifically, the first row shows a clear systematic trend that the size of the hole pocket around the M point reduces significantly, as the intralayer magnetic correlation $\chi_\parallel$ of the Ni-O layer changes from strongly FM (left) to strongly AFM (right) under a fixed AFM interlayer correlation $\chi_\perp$.
Correspondingly, the electron pockets around the A point also shrink their size.
(These electron pockets originate from electron transfer from Ni$^{+}$ to Nd$^{3+}$ in the absence of chemical doping, a phenomenon commonly known as ``self-doping''~\cite{GMZhang}.)
A similar strong reduction of the Fermi pocket is also observed in the second row, when the interlayer correlation changes from FM to AFM while keeping the intralayer correlation $\chi_\parallel$ AFM.
That is, AFM correlation can efficiently reduce the Fermi pockets, or more essentially the carrier density.

\begin{figure}
\centering
\includegraphics[width=1\columnwidth]{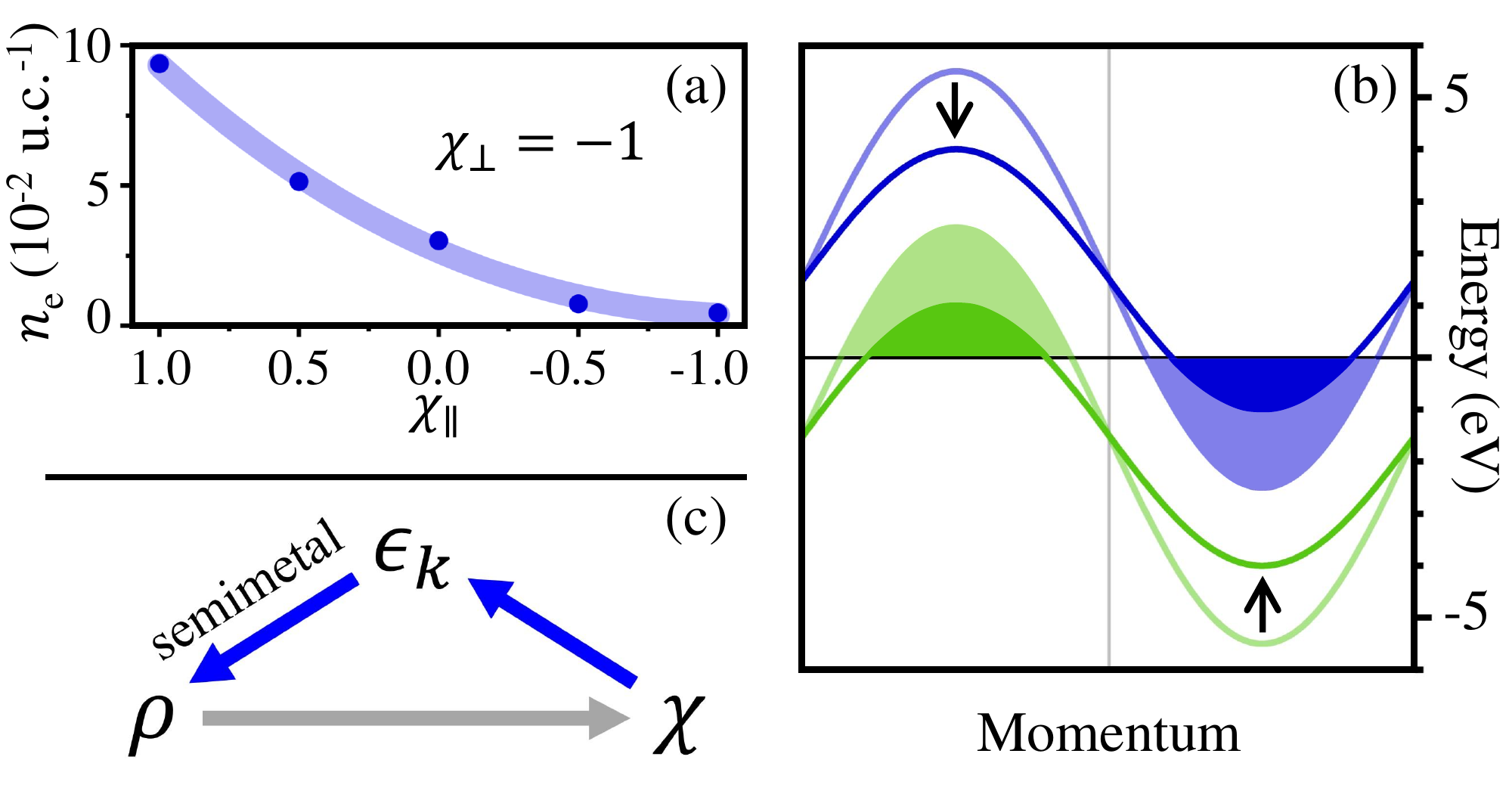}
\vspace{-0.8cm}
\caption{(a) Significant suppression of carrier density under stronger intralayer AFM \textit{noncollinear} correlation $\chi_\parallel$, corresponding to the first row in Fig.~\ref{fig2}.  (b) Illustration of reduced carrier density via weakening of kinetic energy, showing smaller (blue) electron- and (green) hole-pockets.  (c) Illustration of the unique mutual influence between carrier density $\rho$ and interatomic correlation $\chi$ in semimetals, through renormalization of the kinetic energy $\epsilon_k$.  The blue arrows indicate the reverse influence of $\chi$ on $\rho$ found in this study.
}
\vspace{-0.6cm}
\label{fig3}
\end{figure}

This effect of carrier density modulation in correlated semimetals can be quantified by the ensemble-averaged density of electron carrier density (and equivalently that of the hole carriers),
$n_{\rm{e}} = \left \langle \frac{1}{V}\int d^3K \sum_{J\in \left \{ J_{\rm{e}}  \right \}} n_F(\epsilon_{KJ}-\mu)\right \rangle$,
in which the summation involves only those bands contributing to the electron pockets $\left\{J_{\rm{e}} \right\}$, namely those with orbital character $M$ predominantly from Nd and the interstitial ``I'',
$\sum_{M\in \left \{\rm{Nd,I}\right \}}|\braket{KM}{KJ}|^2 > \frac{1}{2}$.
Here $V$ denotes the volume of the system with periodic boundary condition for each configuration, $n_F$ the standard Fermi-Dirac distribution function, and $\mu$ the chemical potential obtained from the \textit{ensemble averaged} one-body spectral function.
Consider the cases in the first row of Fig.~\ref{fig2} as examples, Fig.~\ref{fig3}(a) demonstrates clearly that as the system develops stronger AFM intralayer correlation, the carrier density can be dramatically suppressed in correlated semimetals (by more than \textit{an order of magnitude} in this case.)

We stress that this important effect is from the short-range correlation, instead of the long-range order.
Since a long-range order necessarily implies certain strength of short-range correlation, typical studies incorporating the former unavoidably inherit the impacts of the latter as well.
However, in the presence of long-range fluctuation,
a system with strong short-range correlation does not necessarily host a long-range order.
Our result makes clear that it is really the short-range correlation, \textit{not} the long-range order, that gives rise to the observed density modulation (and many previously reported observations in ordered structures~\cite{Liu2020,Choi2020_1,Choi2020_2}).

Also note that even though our demonstration above focuses on carrier scattering due to magnetic correlation, the microscopic mechanism discussed here is generally applicable to all strong short-range correlation.
Since all short-range charge-, lattice-, or orbital-correlations act to restrict the carrier motion to some degree regardless their microscopic details, they produce a similar effect of carrier density modulation in correlated semimetals.

The above significant effect of short-range correlation on carrier density actually has an intuitive microscopic origin, namely the reduction of effective kinetic energy.
Due to the Pauli principle and other many-body couplings, when moving between neighboring atoms, carriers encounter different degrees of scattering probability: weaker with a similar spin environment and stronger with rapidly varying spin orientation.
This naturally introduces a correlation-dependent reduction of the effective kinetic energy of the carriers and at the long length scale a reduction of their bandwidth.
As illustrated in Fig.~\ref{fig3}(b), in semimetals this would in turn shrink the size of electron and hole Fermi pockets and decrease the corresponding carrier densities.
Altogether, combined with the unique flexibility of carrier density in semimetals, short-range correlation's renormalization of kinetic energy can \textit{reversely} affect the carrier density, as indicated by the blue arrows in Fig.~\ref{fig3}(c).

The first and third rows of Fig.~\ref{fig2} confirm this intuitive picture.
As the in-plane short-range AFM correlation increases ($\chi_\parallel \rightarrow {-1}$), the Ni bandwidth is significantly reduced by the stronger scattering.
This leads to a much smaller carrier density easily observable from the removal of the electron pockets near the $\Gamma$ point.
In comparison, the fourth row verifies that as long as the intralayer correlation is uniform or FM, only very limited damage to the intralayer kinetic energy can be induced by increasing interlayer AFM correlation, as expected from the above picture.

\section{MUTUAL INFLUENCE BETWEEN CARRIER DENSITY AND CORRELATION}
Generally speaking, since carrier density is one of the most effective factors that control the correlation strength [cf. the gray arrow in Fig.~\ref{fig3}(c)], the above-mentioned reverse influence of the latter to the former indicates a unique mutual influence between them.
Such a non-linear feedback illustrated in Fig.~\ref{fig3}(c) is the perfect recipe for sensitive bifurcating behavior, toward either a weakly correlated metal with high carrier density or a strongly correlated system with low density.
For example, in the context of semimetal-semiconductor transition, it should strengthen Mott's proposal~\cite{Mott1968} of a first-order quantum phase transition.

We now reach the answers to our key scientific questions on semimetals concerning (1) the key factor controlling the carrier density variation, and (2) the origin of the carrier density's sensitivity to weak external conditions.
The above-mentioned reverse influence of interatomic correlation on carrier density is the natural candidate that controls the carrier density variation at low energy.
Furthermore, this mechanism is active with a small energy scale ($<$100meV) relevant to the external parameters such as temperature, pressure, and external field.
This relevant energy scale of interatomic correlation, together with the non-linear sensitivity due to its mutual influence with carrier density, explains the observed high sensitivity of carrier density to these weak external parameters.

Indeed, this mechanism offers simple explanations to numerous exotic observations in many correlated semimetals.
For example, it allows intuitively the dramatic reduction of the Hall coefficient (by orders of magnitude below 300 K)~\cite{Li2019, Ghosh2010} in Ni-based and Fe-based unconventional superconductors.
Similarly, it accounts for the significant shrinking of the Fermi pockets in these materials when comparing with standard band theories.
Also, the extreme sensitivity of superconductivity to the substrate and pressure in these materials~\cite{Kawai2018,Ren2021,Wang2022} can also result (or benefit greatly) from the same consideration.

\section{DISCUSSION FOR N\MakeLowercase{d}N\MakeLowercase{i}O$_2$}
Specifically for the prototypical NdNiO$_2$, our study reveals several important effects in the quasiparticle dispersion and lifetime.
Figure~\ref{fig2} shows that the electron pockets around the $\Gamma$ point via band calculations would disappear under strong enough AFM correlation.
Furthermore, the magnetic correlation enhanced scattering is much stronger for the hole carriers in the Ni orbitals, such that their propagation is more diffusive than ballistic.
Correspondingly, the hole carriers can easily lose the quasiparticle nature and become susceptible to non-Fermi liquid physics, such as the strange metal and bad metal behavior~\cite{Li2019}.
Interesting, near momenta points $M$ and $A$, the suppressed kinetic process of the Ni orbital causes them to retract from the Fermi energy, leaving mostly the O orbital at low energy and in turn strengthens the charge-transfer nature.
This reinforces the claim~\cite{lang2021} that in this system the hole carriers reside primarily in the O $p$ orbitals and form Zhang-Rice singlets with the intrinsic holes in Ni.
Our results also clarify that most features found previously in the theoretical band structures of magnetically ordered nickelates actually result from the short-range correlation instead of the long-range orders.
One example is the emergence of a prominent flat band near the chemical potential in the $k_z=\pi$ plane, which can potentially promote various instability~\cite{Choi2020_2} at low temperature.
Finally, the heaviness of the diffusive hole carriers and their low density are both harmful to the superfluid stiffness.
This indicates the need for sufficient doping in establishing a stronger phase coherence, in good agreement with the observed phase diagram.


\section{CONCLUSION}
In summary, we identify a generic low-energy mechanism for the puzzling tunability of carrier density in correlated semimetals.
To incorporate the \textit{interatomic} correlations and their physical impacts that poses a serious technical challenge to the state-of-the-art methods, we develop a DFT-based computational simulation scheme for the one-body propagator under \textit{noncollinear} magnetic correlation \textit{without} long-range order.
Using recently discovered Ni-based unconventional superconductors as an example, we demonstrate significant impacts on the resulting quasiparticle dispersion and lifetime.
Moreover, in contrast to the well-known modulation of correlation via carrier density, in semimetals short-range correlation can \textit{reversely} affect carrier density.
Such mutual influence suggests an enhanced tendency toward bifurcating physical properties of low-energy scale relevant to slight tuning of external parameters such as temperature, pressure, or external field.
This unique feature of correlated semimetals not only provides a natural explanation for the observed exotic tunability of carrier density in many materials, but also suggests routes to functionalize correlated semimetals with richer physical properties and wider scope of application in electronic devices.

\begin{acknowledgments}
This work is supported by the National Natural Science Foundation of China (NSFC) under Grant No.12274287 and No.12042507, and the Innovation Program for Quantum Science and Technology No. 2021ZD0301900.
A portion of this work was conducted at the Center for Nanophase Materials Sciences, which is a DOE Office of Science User Facility.
\end{acknowledgments}

\renewcommand{\theequation}{A\arabic{equation}}
\renewcommand{\thefigure}{A\arabic{figure}}
\renewcommand{\thetable}{A\arabic{table}} 
\setcounter{equation}{0}
\setcounter{figure}{0}
\setcounter{table}{0}

\section{Appendix}
\subsection{\MakeUppercase{Computational details of density functional calculation}}
For this prototypical case, we obtain the most relevant Hilbert space within $\pm10$ eV around the Fermi energy from the spin polarized LDA+$U$ ~\cite{Anisimov1993, Liechtenstein1995} electronic structure of the parent compound NdNiO${_2}$, using the linearized augmented plane wave~\cite{Singh} implementation~\cite{Blaha1990} of the density functional theory (DFT)~\cite{DFT1, DFT2}.
We take from Ref.~\cite{Li2019} the lattice structure with lattice constant $a=b=3.92\mathring{A},c=3.37\mathring{A}$, and the space group $P4/mmm$.
We set $U-J$ = $0.6 - 0.06$ Ry = 7.344 eV for Ni $d$ orbitals,$0.20 - 0.00$ Ry = 2.72 eV,  $0.60 - 0.00$ Ry = 8.16 eV for Nd $d$-,$f$ orbitals separately.

\subsection{\MakeUppercase{Wannier orbitals as configuration independent basis}}
\begin{figure}[h]
	\begin{center}
 	\vspace{-0.3cm}
	\includegraphics[width=0.9\columnwidth]{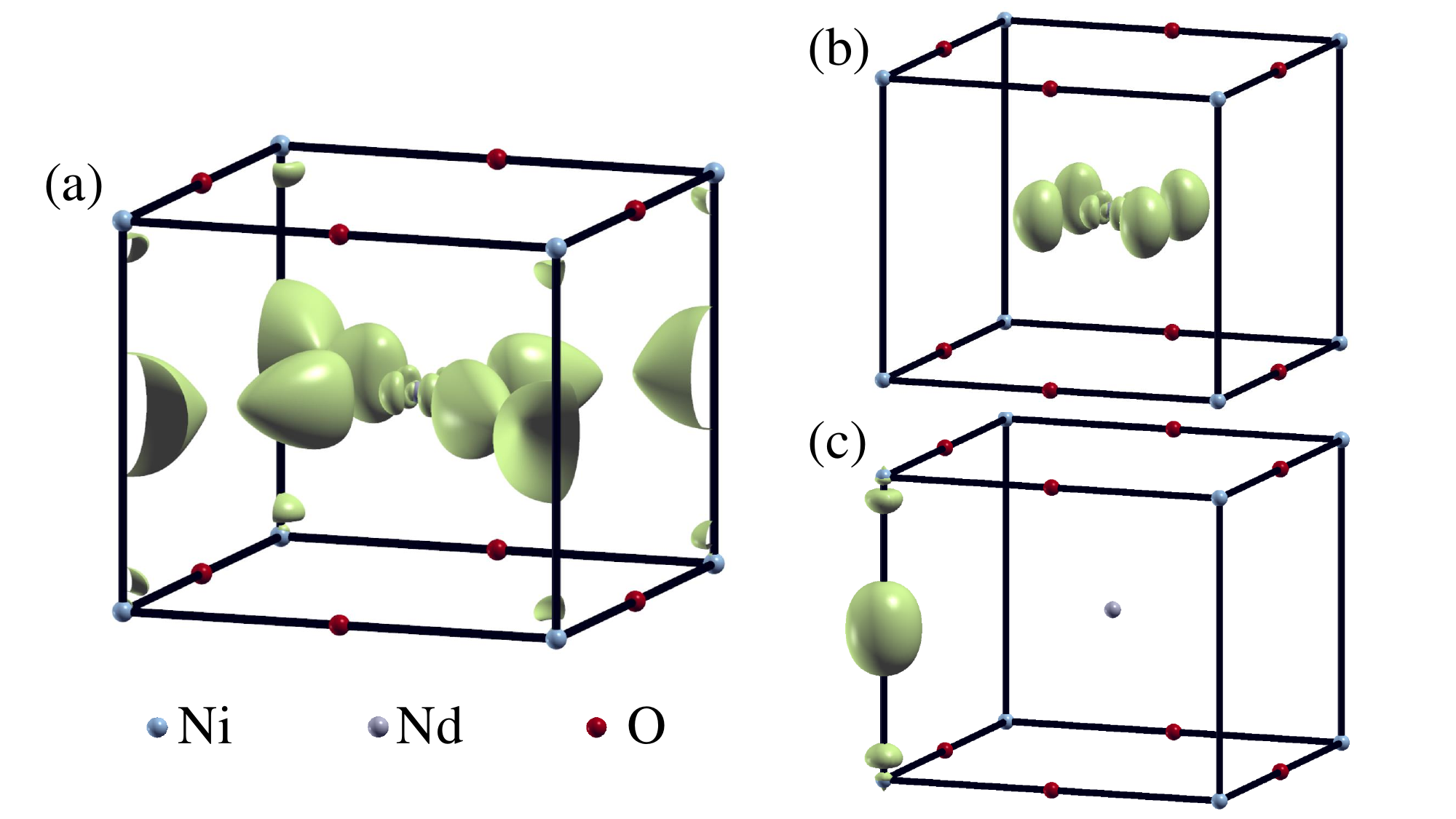}
	\end{center}
	\vspace{-0.5cm}
	\caption{
    Illustration of the Bloch orbital corresponding to the electron pocket at momentum $\mathrm{A}=(\pi,\pi,\pi)$ (a), showing a significant contribution in the interstitial region.  A more complete basis thus should include an interstitial ``I'' orbital (c) in addition to the Nd-$d_{xy}$ Wannier orbital (b).
	}
	\label{figs1}
	\vspace{-0.6cm}
\end{figure}

In disordered systems the orbital hybridization can vary strongly.
It is therefore necessary to construct a more complete set of \textit{charge-active} atomiclike Ni $d$-, O $p$- and Nd $d$-, $f$-Wannier orbitals~\cite{Wei2002, Wei2006, Marzari1997} without downfolding to a smaller low-energy subspace.
In addition, the Bloch orbital corresponding to the electron pocket at momentum $\mathrm{A}=(\pi,\pi,\pi)$ shown in Fig.~\ref{figs1}(a) contains a significant contribution in the empty space above Ni atoms~\cite{Nomura2019}.
It is thus highly beneficial to include an additional interstitial ``I'' orbital [cf. Fig.~\ref{figs1}(c)] in addition to the Nd-$d_{xy}$ Wannier orbital [cf. Fig.~\ref{figs1}(b)].
These symmetry-respecting Wannier orbitals form a nearly complete basis that is atomically local and nearly configuration independent, making them ideal for the ensemble averaging of the resulting one-particle spectral function.

\subsection{\MakeUppercase{Extraction of many-body Hamiltonian}}
We aim at obtaining a realistic SU(2)-symmetric many-body Hamiltonian $H^{\text{eff}}$ in Eq.~(\ref{eq1}).
To ensure a DFT-like level of accuracy, we further demand that when under a similar approximation, our many-body Hamiltonian needs to reproduce the self-consistent DFT Hamiltonian (equivalently the electronic band structure).
Since in LDA+$U$ the strong local electron-electron interaction is approximated in an effective Hartree-Fock manner~\cite{Anisimov1997}, a proper connection can be naturally made on the Wannier states basis by matching the self-consistent Hartree-Fock solution of our SU(2)-symmetric $H^{\text{eff}}$ with the self-consistent $H^{\text{LDA}+U}$~\cite{Wei2006}.
The requirement that the Hartree-Fock approximation of $H^{\text{eff}}$ needs to reproduce the LDA+$U$ band structure within the subspace of the active orbitals results in a rather unique set of parameters in $H^{\text{eff}}$~\cite{Wei2006}.

Specifically, since we demand that the self-consistent Hartree-Fock solution reproduce the self-consistent solution of the LDA+$U$ solution, the corresponding density matrix $\rho_{i\nu\nu^\prime}$ must be identical in both cases when represented in the same set of Wannier orbitals.
One can thus take $\rho_{i\nu\nu^\prime}$ directly from the self-consistent LDA+$U$ solution in the Wannier basis.
Furthermore, if one assumes that the structure of $U_{m m^{\prime\prime}m^\prime m^{\prime\prime\prime} }$ follows the Slater integral~\cite{Slater,Liechtenstein1995}, the entire $U_{m m^{\prime\prime}m^\prime m^{\prime\prime\prime}}$ can be fixed by just two parameters $U_{\text{eff}}$ and $J_{\text{eff}}$.
Combining $\rho_{i\nu\nu^\prime}$ and $U_{m m^{\prime\prime}m^\prime m^{\prime\prime\prime} }$, the effective Hartree-Fock potential $V^\text{HF}$ can then obtained.

In addition, orbital dependent $J_m$ in $H^{\text{eff}}$ can be straightforwardly obtained from the local site energy difference of Nd $d$ orbitals between spin up and spin down, when the $\frac{3}{2}$ spin $\mathbf{S}_i^f$ of the $f$ orbitals are all set to be along the spin up direction.
Finally, the remaining intraatomic hopping parameters $t_{ii mm^\prime}$ and site energy $\epsilon_m$ are then obtained by subtracting $V^\text{HF}$ from the intraatomic part of $H^{\text{LDA}+U}$.

A simple criterion to check the quality of the resulting $H^{\text{eff}}$ (or the accuracy of the chosen $U_{\text{eff}}$ and $J_{\text{eff}}$) is the degree of the spin independence of the intraatomic $t_{ii mm^\prime}$.
A reasonable value of $U_{\text{eff}}$ and $J_{\text{eff}}$ should encode all the spin dependence of the self-consistent $H^\text{HF}$.
Therefore, a non-negligible spin dependence of the resulting $t_{ii mm^\prime}$ or site energy $\epsilon_m$ indicates clearly a need to improve the value of $U_{\text{eff}}$ and $J_{\text{eff}}$.
In practice, we find this criterion quite sufficient to pin down a rather narrow range of acceptable values of $U_{\text{eff}}$ and $J_{\text{eff}}$.
This then allows us to fix all the intraatomic parameters in $H^{\text{eff}}$.

On the other hand, since the LDA+$U$ only included atomically local Hartree-Fock approximation, the interatomic hopping parameters $t_{ii^\prime mm^\prime}$ are unaffected in the approximation.
We therefore can simply take it from the spin averaged $H^{\text{LDA}+U}$.
The next section lists some of the leading parameters.

The above procedure, if performed properly, should generate a SU(2)-symmetric many-body Hamiltonian $H^{\text{eff}}$ that can be further studied with any many-body solver one prefers, not just the Hartree-Fock approximation.
Furthermore, since the Hamiltonian does not require or guarantee a magnetic order, it can in principle be applied to metallic systems (or spin liquid insulators) that do not contain long-range order but show signs of the existence of local moments---for example, Curie-Weiss susceptibility at high-temperature.

\begin{figure}[h]
	\begin{center}
	\includegraphics[width=1\columnwidth]{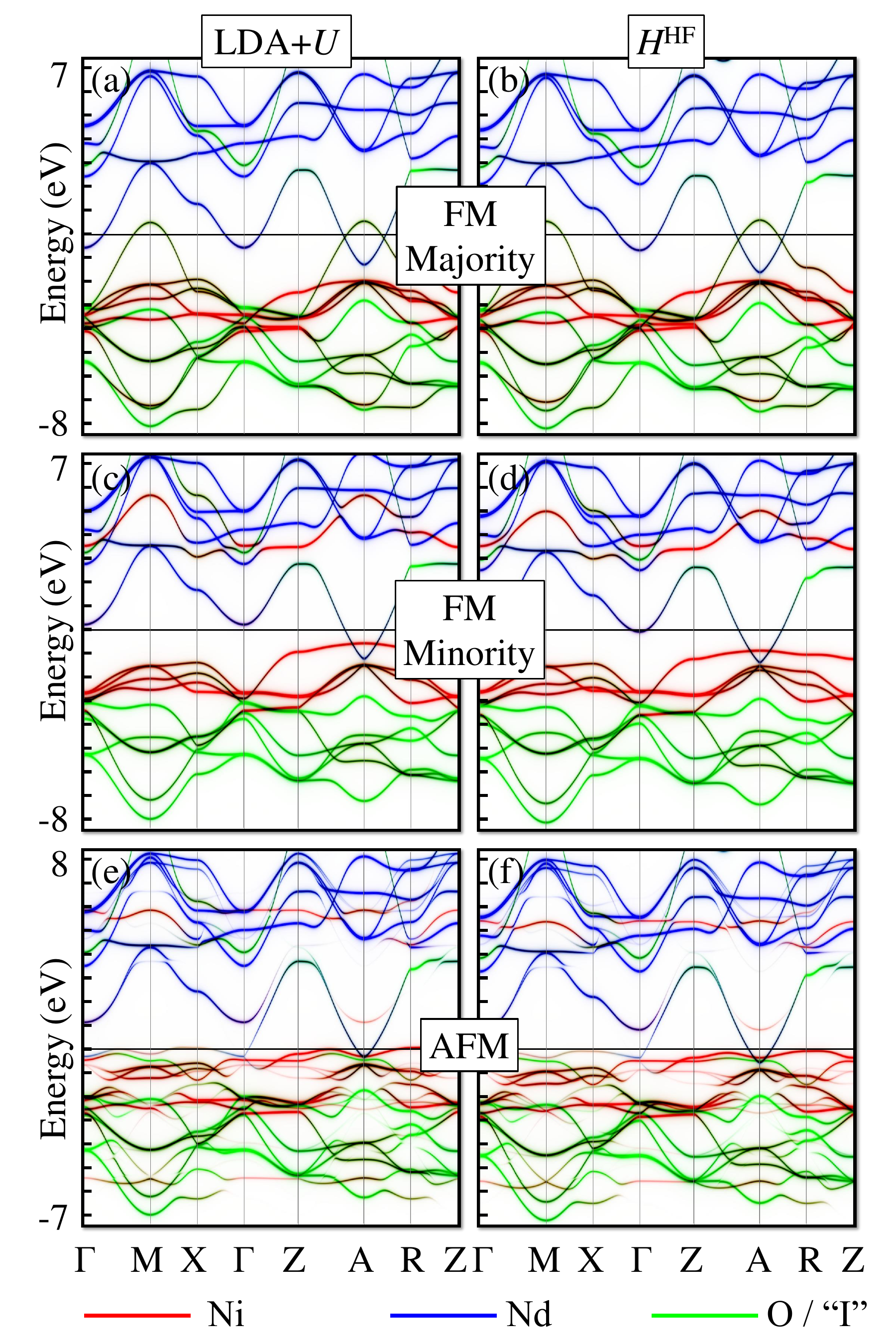}
	\end{center}
	\vspace{-0.6cm}
	\caption{
	Confirmation of the quality of $H^{^\mathbf{eff}}$ by comparing the electronic band structures with those from LDA+$U$ calculations under different magnetic configurations: (a)(b) spin majority of ferromagnetic order, (c)(d) spin minority of ferromagnetic order,  and (e)(f) antiferromagnetic order.
	The excellent agreement establishes a robust foundation for configurations with noncollinear spin correlations without long-range order.
	For clarify, Nd $f$ bands are not shown (see the main text).
	}
	\label{figs2}
	\vspace{-0.3cm}
\end{figure}

For the purpose of this work, in which we study the impact of noncollinear spin correlations, we verify that $H^{\text{eff}}$ is able to reproduce the $H^{\text{LDA}+U}$ solution under various magnetic orders.
Figure~\ref{figs2} demonstrates an excellent correspondence between our resulting band structure under the same Hartree-Fock treatment of local interactions and the LDA+$U$ band structure within the subspace of the selected orbitals.
Notice especially that the agreement occurs under both ferromagnetic and antiferromagnetic order with the \emph{same} set of parameters, despite the significantly different band structure under these two orders.
These results thus establish the high quality of these parameters and the validity of our effective Hamiltonian in describing various magnetic structures, including the paramagnetic parent compound.

\subsection{\MakeUppercase{Leading parameters for $H^{\text{eff}}$}}
\begin{table}[t]
    \caption{Leading parameters $\epsilon_m$ and $J_m$ in unit of eV.}
    \begin{ruledtabular}
        \begin{tabular}{llllll}
Nd & $d_{3z^2-r^2}$ & $d_{x^2-y^2}$ & $d_{yz}$ & $d_{xz}$ & $d_{xy}$ \\ \hline
$J_m$  & 0.272 & 0.379  & 0.268 & 0.268 & 0.217 \\
$\epsilon_m$ & 4.12 & 5.43  & 5.779 & 5.779 & 4.045 \\ \hline \hline
Ni & $d_{3z^2-r^2}$ & $d_{x^2-y^2}$ & $d_{yz}$ & $d_{xz}$ & $d_{xy}$ \\ \hline
$\epsilon_m$ & -55.448 & -55.101  & -55.616 & -55.616 & -55.284 \\\hline\hline
O & $p_{x}$ & $p_{y}$ & $p_{z}$ & ``I'' \\ \hline 
$\epsilon_m$ & -3.488 & -2.542  & -2.602 & 4.93 \\
\end{tabular}
    \end{ruledtabular}
    \label{tabs1}
\end{table}
Table \ref{tabs1} lists some of the leading coefficients $\epsilon_m$ and $J_m$ of the resulting SU(2) symmetric many-body Hamiltonian.
The full interaction kernel $U_{m m^{\prime\prime} m^\prime m^{\prime\prime\prime}}$ is approximated by the Slater integral~\cite{Slater,Liechtenstein1995}, with $U_{\text{eff}} = 7.385$ eV and $J_{\text{eff}}=1.197$ eV obtained from the above procedure.
The leading terms of the resulting hopping parameters $t_{ii^\prime mm ^\prime}$ are $1.321$ eV between Ni $d_{x^2-y^2}$ and O $p_{x}$, $1.158$ eV between Nd $d_{xz}$/$d_{yz}$ and O $p_{z}$, and $1.024$ eV between Nd $d_{xy}$ and ``I'' orbitals.
The full $t_{ii^\prime mm ^\prime}$ parameters are available upon request.

\subsection{\MakeUppercase{Implementation of noncollinear magnetic configuration}}
\begin{figure}[h]
	\begin{center}
	\includegraphics[width=0.85\columnwidth]{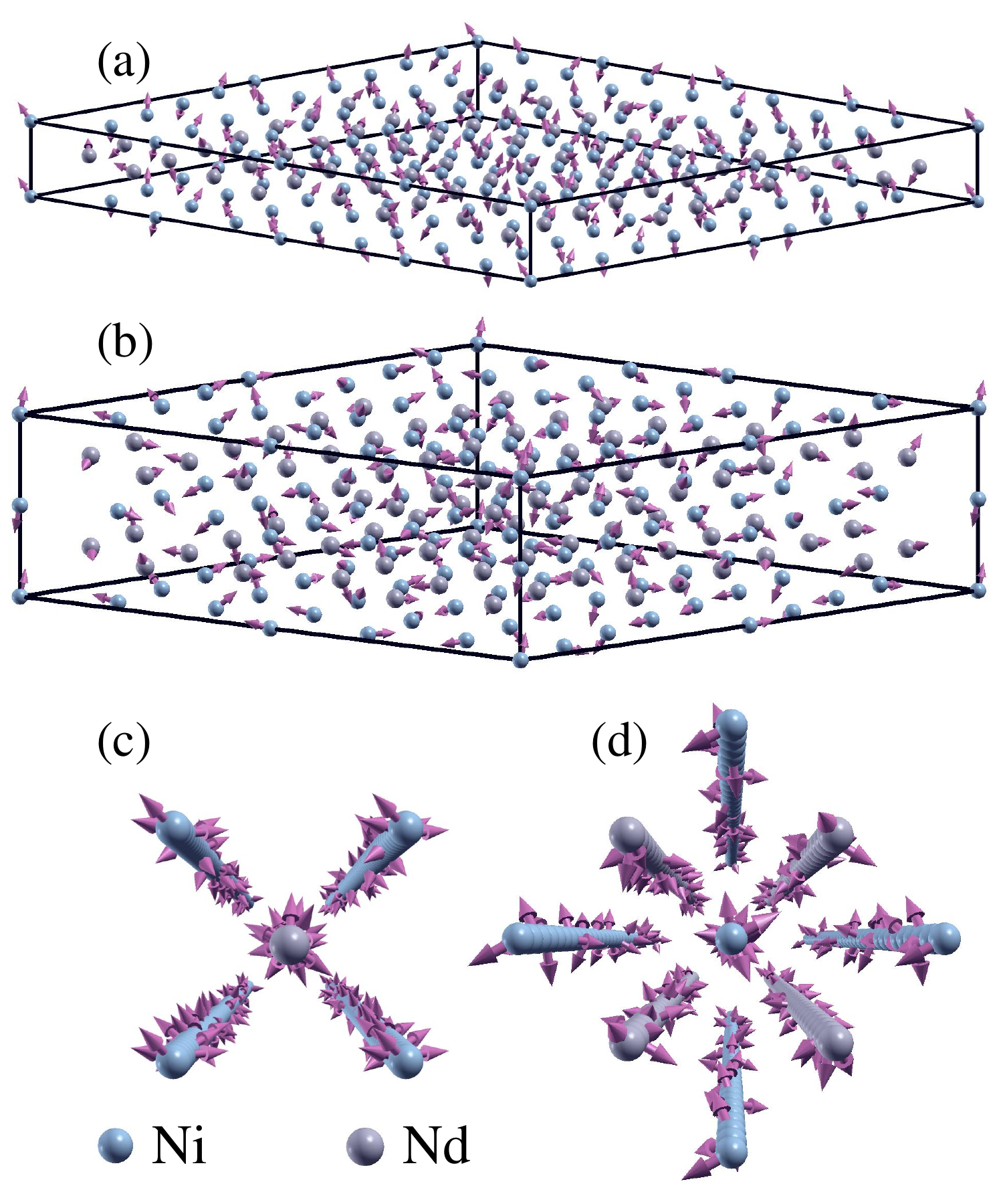}
	\end{center}
	\vspace{-0.2cm}
	\caption{Examples of supercells from categories: (a)$\chi_\parallel=$ 1,$\chi_\perp\sim$ -0.5, (b)$\chi_\parallel=$ -1,$\chi_\perp\sim$ 0, (c)$\chi_\perp=$ 1,$\chi_\parallel\sim$ -0.5, and (d)$\chi_\perp=$ -1,$\chi_\parallel\sim$ 0.  Purple arrows represent the spin directions of Nd and Ni atoms.
	}
	\label{figs3}
	\vspace{-0.2cm}
\end{figure}

\begin{figure}[h]
	\begin{center}	\includegraphics[width=1\columnwidth]{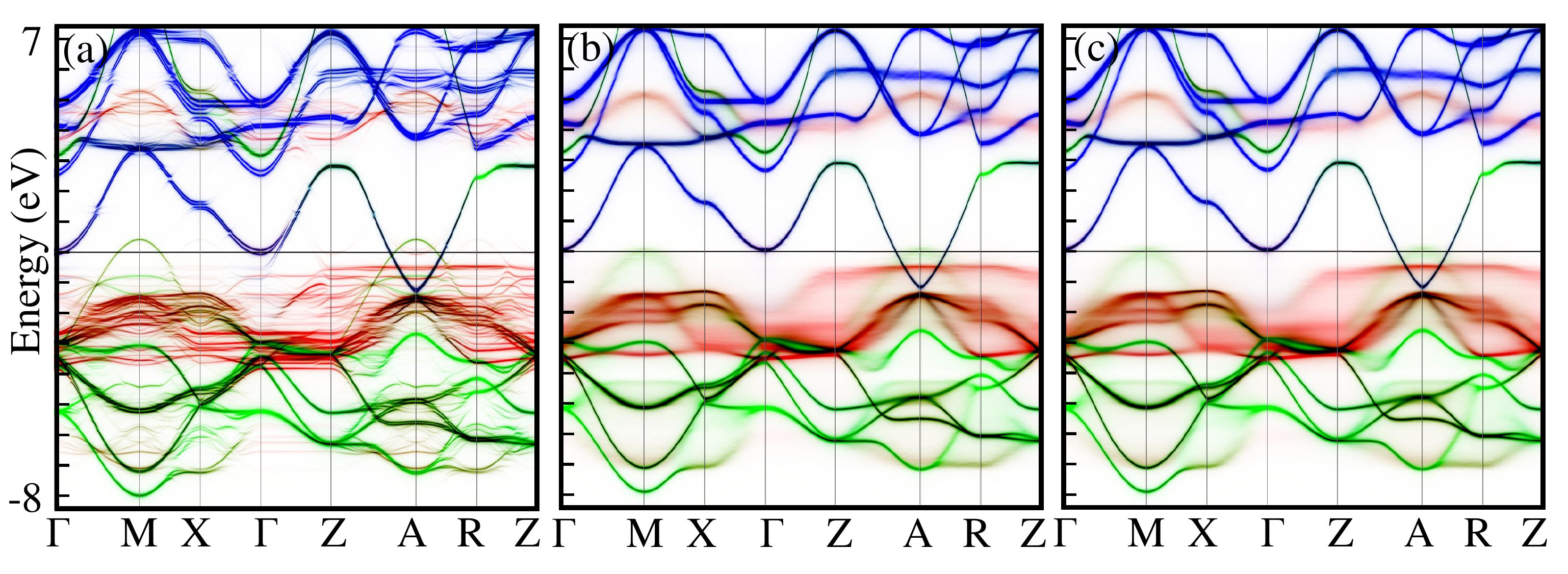}
	\end{center}
	\vspace{-0.6cm}
	\caption{
    Band structure corresponding to (a) $2\cross 2 \cross 2$ supercell, (b) ensemble average over 10 large supercells, and (c) ensemble average over 22 large supercells.
	}
	\label{figs4}
	\vspace{-0.2cm}
\end{figure}

In this study, the fluctuation of noncollinear spin directions is incorporated in our simulation by averaging a large number of configurations with similar nearest short-range correlations $\chi = \sum_{\langle i, i^\prime\rangle}\mathbf{S}_{i} \cdot \mathbf{S}_{i^\prime}$, 
but with negligible long-range order $\mathbf{M}_{\mathbf{q}} = \sum_{i}\mathbf{S}_{i} e^{i\mathbf{q}\cdot \mathbf{x}_i}$.
Obviously, a larger supercell is necessary in the search for configurations of various short-range correlation, but without long-range order.
Furthermore, averaging over large supercells (cf. examples in Fig.\ref{figs3}) with various sizes, shapes and orientations is an efficient way to avoid fictitious gap openings and shadow band foldings related to the artificial new spatial periodicity.
(Figure~\ref{figs4}(a) gives an example showing artificial shadow bands and gap opening at \textit{particular} momenta resulting from the periodicity of a single configuration with a rather small supercell.)

The enforcement of the spin directions according to the proposed magnetic configurations can be easily achieved within the Hartree-Fock approximation by constraining the atomically local one-body density matrix $\rho_{i\nu\nu^\prime}$ to be diagonal in the spin channel $\nu=\nu^\prime$ along the assigned spin direction in every self-consistent cycle.
Specifically, using the Euler angle we represent (rotate) $\rho_{i\nu\nu^\prime}$ in a local spin basis whose $z$ axis is along the assigned spin direction.
We then zero out the off-diagonal elements of $\rho_{i\nu\nu^\prime}$ in this local basis, and then rotate the representation back to the global one with the $z$ axis of the spin along that of the lattice.
This constrained $\rho_{i\nu\nu^\prime}$ is then combined with  $U_{m m^{\prime\prime} m^\prime m^{\prime\prime\prime}}$ in Eq.~(\ref{eq1}) to evaluate the effective orbital-dependent potential within the Hartree-Fock approximation.
Naturally, the same constraint needs to be applied in each iteration of the self-consistent cycle, until a spin density is converged.

Convergence of our results is illustrated in Fig.~\ref{figs4}(c).
Upon doubling the number of distinct configurations,  Fig.~\ref{figs4}(c) shows essential features practically indistinguishable from Fig.~\ref{figs4}(b) as the paramagnetic results shown in Fig.1(b) of the manuscript.

\subsection{\MakeUppercase{Formal theoretical foundation of the simulation}}
\begin{figure}[h]
	\begin{center}
	\includegraphics[width=0.8\columnwidth]{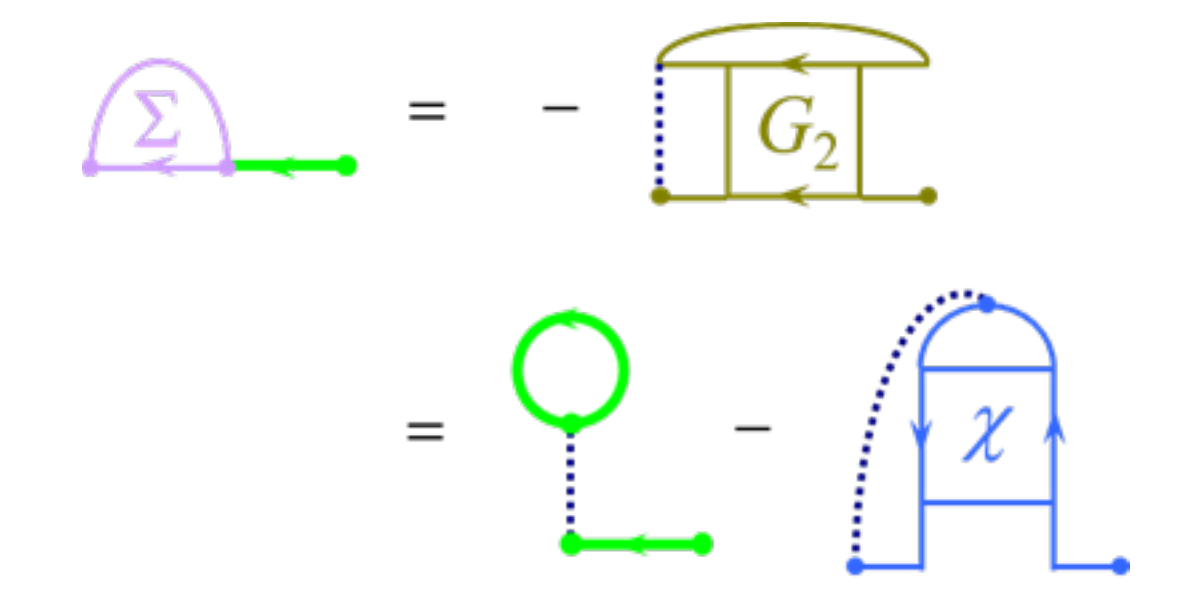}
	\end{center}
	\vspace{-0.6cm}
	\caption{
    Exact one-body self-energy $\Sigma$ is only sensitive directly to the two-body correlation functions.  Here, green lines and dotted lines represent the one-body Green's function and the two-body interaction.  $G_2$ and $\chi$ denote the two-body Green's function and the two-body correlation function.
    }
	\label{figs5}
	\vspace{-0.2cm}
\end{figure}

At the formal conceptual level, our simulation is conceptually clean in formulation.
This is because the main physics of interest in this study is through the strong renormalization of the kinetic energy present in the one-body propagator, which is only sensitive directly to the two-body correlation functions even in the exact many-body formulation.
As shown in the Figure~\ref{figs5}, the exact self-energy $\Sigma$ of the one-body Green's function that encapsulates the band dispersion can be obtained fully with two-body correlation functions, through which effects of higher order N-body correlations are all included.
Therefore, regardless of the underlying quantum or classical origins of the two-body correlation functions and their approximate leading contributions, our simulation is conceptually clean in formulation.

\subsection{\MakeUppercase{Consideration and comparison with the LDA+DMFT approach}}
Our study aims at illustrating a mechanism for sensitive tunablility of material properties unique in correlated semimetals often observed in experiment under weak (low-energy) external control such as temperature, pressure, or external field.
The low-energy sensitivity renders the high-energy (10 eV-scale) intraatomic physics inadequate.
Instead, we seek physics originating from lower-energy (100 meV or lower) physics, specifically the interatomic correlation.

To this end, the state-of-the-art dynamical mean-field treatment~\cite{Petocchi2020,Ryee2020,Kitatani2020,Lechermann2021} is not suitable, since it can only incorporate the high-energy intraatomic correlation but is incapable of including the multiple scattering resulting from interatomic correlation.
(The $GW$ calculation~\cite{Hirayama2022,Petocchi2020} is even less suitable since it includes only screening of long-range charge fluctuation but incorporates poorly the short-range correlation.)
Specifically, since the key physical energy scales of $GW$ (the plasmon frequency) and DMFT (intraatomic repulsion) are both of the order of 10 eV, roughly two to three orders of magnitude larger than the room temperature, they cannot possibly be directly related to the versatile temperature-dependence of interest in this manuscript.  
The results of these methods would instead show negligible temperature dependence of carrier density at T$\sim$300K and below.
On the other hand, since the single-hole valence of Ni $d^9$ does not support a strong intraatomic many-body multiplets, the local Hartree-Fock treatment should be a reasonable approach, especially considering the need for very large noncollinear configurations in our simulation.
Given that the state-of-the-art results in the literature are primarily from the LDA+DMFT calculations, we compare below our result of the paramagnetic case with that of LDA+DMFT calculation to illustrate the important features from our new capability.

\begin{figure}[h]
	\begin{center}
	\includegraphics[width=1\columnwidth]{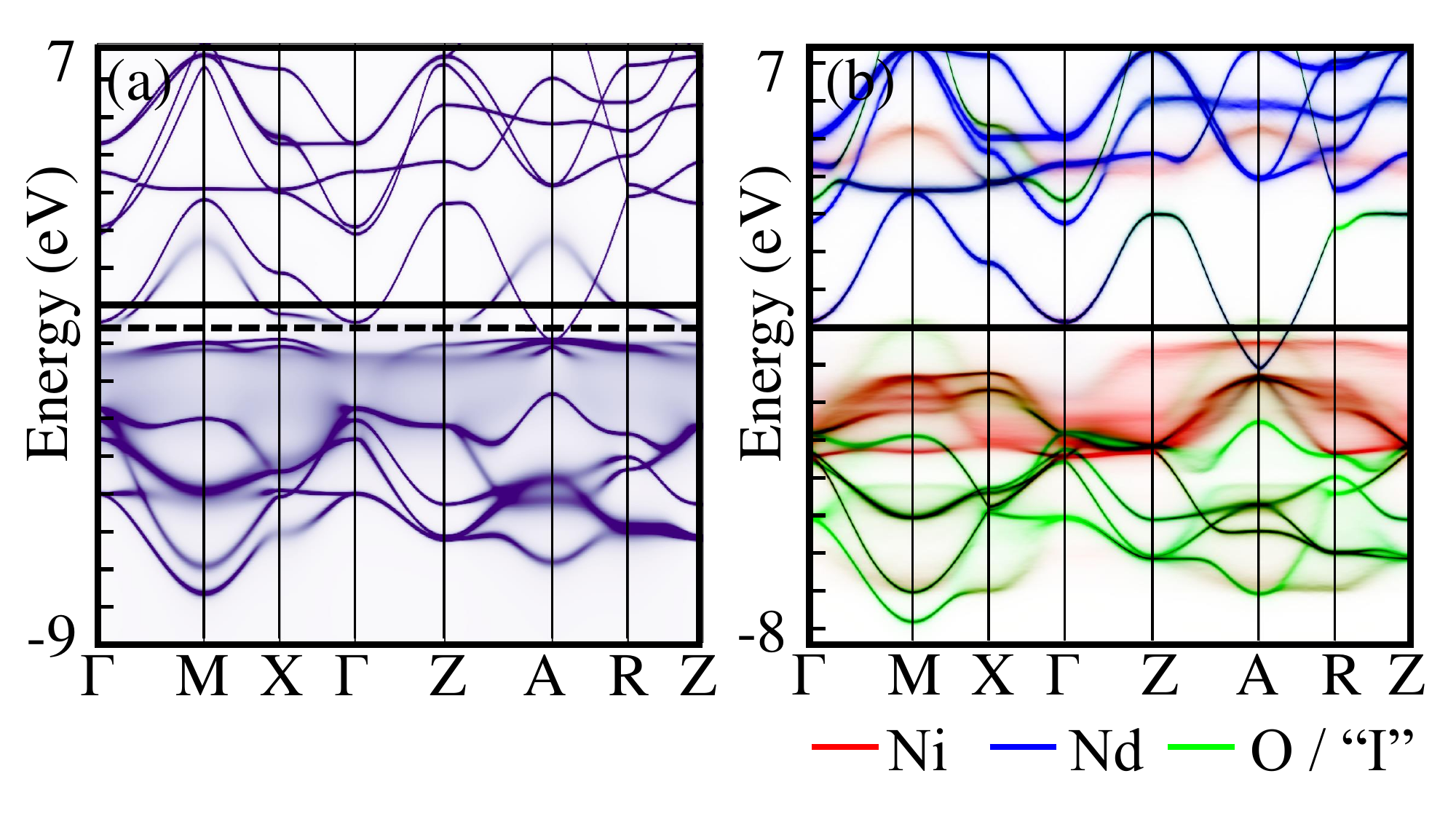}
	\end{center}
	\vspace{-0.6cm}
	\caption{
    Comparison of resulting one-body spectral function for the paramagnetic case from (a) charge self-consistent LDA+DMFT calculation and (b) self-consistent noncollinear Hartree-Fock simulation, with identical lattice parameters.
	}
	\label{figs6}
	\vspace{-0.2cm}
\end{figure}

Figure~\ref{figs6}(a) shows our charge self-consistent LDA + DMFT calculations using the EDMFTF package\cite{Kotliar2006, Haule2007} with all parameters identical to the self-consistent noncollinear magnetic calculation [Fig.~\ref{figs6}(b)] reported in Fig.1(e) of the manuscript.
Clearly, the ``uncorrelate'' O $p$ band and Nd $d$ bands are very similar between these two calculations.  
While the correlated Ni $d$ orbitals show a similar smearing effect as a result of many-body scattering, the general renormalization in these two calculations are quite different.
The main effects of intraatomic correlation in DMFT in Fig.~\ref{figs6}(a) are to 
(1) compress the Ni $d$ bands from LDA to a narrower energy range, and 
(2) to produce a large decay (imaginary part of the self-energy) of the Ni $d$ orbital in some frequency range (around [-3,-1] eV) in which the $d$ bands are barely recognizable (loss of quasiparticle nature). 
Indirectly on the O $p$ bands, the first effect also slightly weakens the bandwidth (due to a larger $d$-$p$ energy separation), while the second effect smears slightly some of the $p$ bands due to hybridization with Ni $d$ orbitals.

In comparison, the effects of interatomic short-range correlation in our paramagnetic calculation (with zero average interatomic correlation) shown in Fig.\ref{figs6}(b) are very different in nature.  
While the kinetic energy of Ni $d$ orbitals is also reduced (for a different physical reason), the smearing effect is however $k$-dependent.  For example, around -2 eV the band between $\Gamma$ and Z point becomes very smeared, while at the same energy the bands around the M point remain well defined.  
This is because the self-energy containing interatomic correlation is beyond atomically local and therefore acquires $k$-dependence absent in the local self-energy in DMFT.

Another obvious difference is the appearance of the upper Hubbard band around [4,6] eV in Fig.\ref{figs6}(b), which is absent in Fig.\ref{figs6}(a) and instead shows up around [-1,2] eV as a heavier version of the LDA band. 
Now, consider Hamiltonians with only large intraatomic interaction $U$ but without interatomic interaction, such as Eq.~(\ref{eq1}). 
Since there is no easy way to screen the large local charging energy $U$ upon adding or removing an electron in the Ni $d$ orbitals, the one-body spectral function corresponding to such addition and removal process must contain a $U$-scale splitting of the occupied and unoccupied $d$ bands (sometimes referred to as the lower Hubbard bands and upper Hubbard bands.)  
Therefore, disregarding whether Fig.\ref{figs6}(a) or Fig.\ref{figs6}(b) might resemble the real material better, the result in Fig.\ref{figs6}(b) clearly captures more faithfully the physics of the effective model Hamiltonian Eq.~(\ref{eq1}) of the manuscript than the DMFT treatment.

\bibliography{main.bib}

\end{document}